\documentclass[useAMS,usenatbib]{mn2e}
\usepackage{graphicx}
\usepackage{natbib}

\def\eso{European Southern Observatory, Karl-Schwarzschild-Strasse 2, 85748 Garching bei M{\"u}nchen, DE}
\def\vic{Department of Physics \& Astronomy, University of Victoria, 3800 Finnerty Road, Victoria BC, Canada V8P 5C2}

\title[How to get any masses from inadequate datasets]{Measured and found wanting: reconciling mass estimates of ultra-diffuse galaxies}
\author[Laporte et al.]{
\parbox[t]{\textwidth}{ Chervin F. P. Laporte$^{1}$\thanks{$\!\!$ CITA National Fellow e-mail:cfpl@uvic.ca}, Adriano Agnello$^{2}$\thanks{$\!\!$email aagnello@eso.org, ORCID 0000-0001-9775-0331}, Julio F. Navarro$^{1}$
}\\
  \medskip\\
  $^1$\vic\\
  $^2$\eso\\
\\
}
\begin{document}
\date{}
\pagerange{\pageref{firstpage}--\pageref{lastpage}} \pubyear{2011}
\maketitle
\label{firstpage}
\begin{abstract}

The virial masses of ultra-diffuse galaxies (UDGs) have been estimated using the kinematics and abundance of their globular cluster populations, leading to disparate results. Some studies conclude that UDGs reside in massive dark matter halos while \cite{PvD18}, controversially, argue for the existence of UDGs with no dark matter at all. Here we show that these results arise because the uncertainties of these mass estimates have been substantially underestimated. Indeed, applying the same procedure to the well-studied Fornax dwarf spheroidal would conclude that it has an `overmassive' dark halo or, alternatively, that it lacks dark matter. We corroborate our argument with self-consistent mocks of tracers in cosmological halos, showing that masses from samples with $5 < N < 10$ kinematic tracers (assuming no measurement errors) are uncertain by at least an order of magnitude. Finally, we estimate masses of UDGs with HST imaging in Coma and show that their recent mass measurements (with adequate uncertainties) are in agreement with those of other dwarfs, such as Fornax. We also provide bias and scatter factors for a range of sample sizes and measurement errors, of wider applicability.

\end{abstract}
                       
\section{Introduction}

Deep and wide-field surveys with fast optics \citep[e.g.][]{delgado08, Pvd14} are leading the exploration of the low-surface-brightness (LSB) frontier of galaxies and their associated halos, extending early studies of galaxy halos and LSB galaxies \citep[e.g.][]{Sandage1984, Turner1993, Dalcanton97, Shang98}. Observations in the Coma cluster have resulted in what has been reported as a new class of LSB galaxies, with kpc-scale half-light radii (some with $R_{\rm{eff}}\approx3$ kpc, comparable to the Milky Way) and $\mu_{\rm{e},r} > 24.5 \,\rm{mag/arcsec^2}$, dubbed ``ultra-diffuse galaxies'' (UDGs; although see Turner et al. 1993 for earlier examples). Galaxies with these properties have since been detected in different environments, in both groups and in the field \citep[e.g.][and references therein]{vdb17, Roman17}. In the Local Group, some dwarf galaxies are even more striking than UDGs (Collins et al. in prep) in terms of surface brightness and sizes, as is the case of Crater 2 \citep{Torrealba16} and And XIX \citep{Martin2016}.

Theoretically, there is no shortage of models for UDGs. \citet{Dalcanton1997} and \citet{amo16} described their population properties as the high-spin and small-mass part of the pre-infall halo population. Numerical simulations suggest that UDGs can form in groups and clusters \citep{dicintio17,chan17} and so they may not be so ``special'' after all. In fact, the interplay between feedback and angular momentum was already explored in order to explain the size distribution of disk galaxies decades ago \citep[e.g.][]{dekel86, Navarro94}. Still, direct probes of halo masses are needed to definitely bridge theory and observations.\\
\indent Globular clusters (GCs) are bright enough that they can be easily detected out to the Coma cluster, thus providing estimates of the host halo masses of faint galaxies from their GC specific frequency \citep{hudson14}. GC counts in Coma suggested that UDGs can have `overmassive' halos, up to $\approx10^{12}\,\rm{M_{\odot}}$ \citep{PvD16}. A follow-up study \citep{Amorisco2018}, on a sample of 54 Coma cluster UDGs with HST imaging data, has shown that they are mostly `normal' dwarf galaxies, with\footnote{$M_{200}$
 being the mass within a radius where the mean inner density is 200 times the critical density of the Universe} $M_{200}\approx 10^9-10^{11}\,\rm{M_{\odot}},$ and that their claimed GC abundances are widely uncertain\footnote{This is in part due to difficulties in estimating membership since contamination effects are non-negligible in Coma.}.

At the opposite end, \citet[hereafter vD18]{PvD18} claimed the discovery of a galaxy lacking DM, NGC~1052-DF2, based on an unusually small velocity dispersion, albeit based on only 10 GCs\footnote{They also argued that their measurement was strong enough evidence to challenge alternative theories of gravity, but it was soon shown by \cite{famaey18} that their study failed to appreciate subtle complications in the MONDian regime.}. This is not the first dwarf with unusually low velocity dispersion, but other teams have been more cautious when interpreting their dark matter content, due mainly to the large uncertainties expected from such small numbers of tracers \citep{toloba18}.  Indeed, the dark matter content estimation depends on a number of procedural steps, each of which is subject to substantial uncertainty:
\begin{itemize}
\item A determination of the true line-of-sight velocity dispersion given the observed velocities and their errors.
\item A conversion of the measured velocity dispersion to mass at a given radius of the tracer (or outer radius depending on the choice of the mass estimator).
\item Extrapolation of the mass within the given radius (e.g. half-light radius) to the virial radius in the case where a virial mass estimation is to be performed.
\end{itemize}

Here we show how some of the recent claims about UDGs, such as NGC1052-DF2, may be uncertain, driven by extrapolation and small number statistics. The same approach, when applied to the nearby Fornax dSph, would lead to inaccurate and even contradictory conclusions: 1) that Fornax lacks DM, and/or 2) that Fornax has a DM halo of $M_{200}\sim2.5\times10^{10}\, \rm{M_\odot}$, about five times larger than inferred from stellar-kinematic data. This Letter is organized as follows. In Section 2 we apply the same approach as vD18 to the Fornax dSph, and provide a re-assessed velocity dispersion of NGC1052-DF2 from the data of vD18. We provide a general evaluation of bias and scatter in $\sigma^2$ from small sample size and velocity measurement errors. We also study the systematics of low $N$ tracers in mass estimators through the sampling of equilibrium distribution functions to corroborate our arguments. We put Fornax and NGC1052-DF2 in the wider context of UDG mass sequences in Section 3, and summarize our findings in Section 4.
\section{Uncertain mass measurements}
Here, we show how nearby dwarf galaxies might be interpreted as `overmassive' or DM-free if the same approach used for UDGs is followed. The simplest example is the Fornax dSph, whose mass within the half-light radius from extended measurements of its stellar populations is very well constrained.
\begin{figure}
\includegraphics[width=0.5\textwidth,trim=0mm 5mm 0mm 20mm, clip]{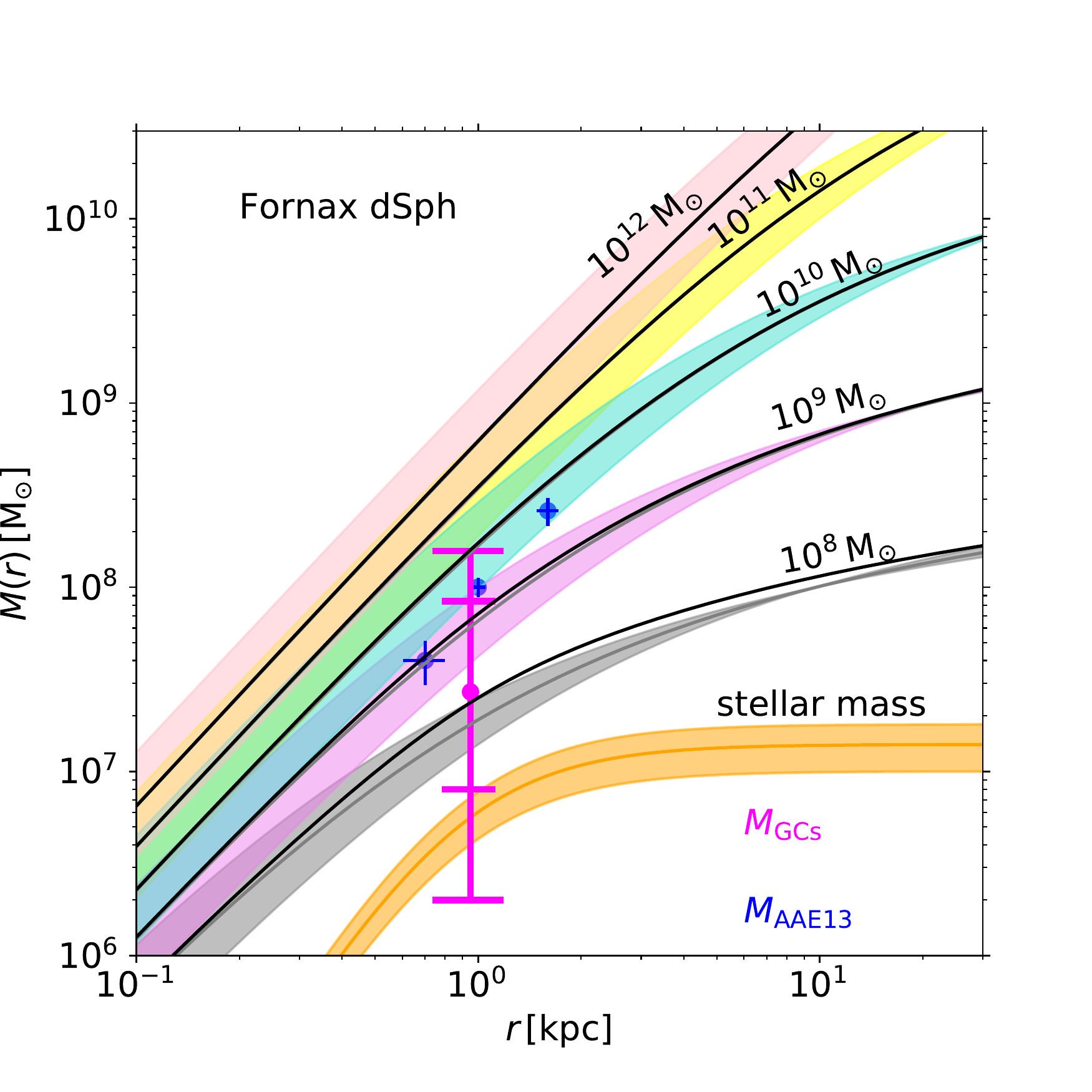}\\
\caption{Mass profile of NFW haloes and constraints on that of the Fornax dwarf spheroidal galaxy. The blue points are constraints on the Fornax mass profile from stellar kinematics \citep{aae13}. The magenta symbol indicates the constraints on Fornax from the velocity dispersion of the system of 4 member globular clusters.  The error bars are 1$\sigma$  and 2$\sigma$  equivalents (i.e. encompassing 68 and 95 per cent confidence). The grey lines indicate NFW dark matter halo mass profiles for integer $\rm{log}(M_{\rm 200})$ and the shaded regions around them indicate 95 per cent confidence on these halo mass profiles given the uncertainties on the concentration - mass relation. The orange line surrounded by the light orange shaded region highlights the the stellar mass profile and its 95 per cent confidence interval. Finally, the black lines indicate the total mass profile (dark matter + stars). The mass inferred from the GCs of the Fornax dSph, if correct, would make it `overmassive' (from GC abundance), or `lacking DM' (lower quantiles from GC kinematics) or `just right' (upper quantiles from GC kinematics). Its stellar populations give a more accurate mass (e.g. Amorisco et al. 2013), albeit with $\approx0.3$~dex uncertainties.}
\label{fig:fornax}
\end{figure}

\subsection{The Fornax dSph: an overmassive halo without dark matter?}

Fornax has a stellar luminosity of $L_{V}= (1.4\pm0.4)\times 10^7 \,\rm{M_{\odot}}$, a metal-poor population with $R_{\rm{eff}}\sim 0.9\,\rm{kpc},$ and 5 known globular clusters associated with it \citep[see Table 1 of ][and references therein]{Cole2011}. Line-of-sight velocities for thousands of red-giant-branch stars have been measured \citep{walker09}. While the exact shape of the DM density profile in dSphs is a topic of debate \citep{Walker2011, Agnello2012, Laporte2013b}, {\it all} agree that the mass within the half-light radius is well constrained \citep[e.g. discussions in][]{walker09,Wolf2010}, and that this dwarf is dark-matter dominated with dynamical mass $M(R_{\rm eff}) \sim 5.3\times10^{7}\,M_{\rm \odot}$ enclosed within a 3D radius from the centre equal to $R_{\rm eff} \sim 668 \,\rm{pc}$.

Let us consider a back-of-the-envelope calculation for the mass of Fornax using solely its GCs. Only 4 of the GCs have a measured line-of-sight velocity. Half of them are contained within a (projected) radius of $0.95\pm0.53 \,\rm{kpc}.$
 The most robust approach for estimating $\sigma_{\rm GC}$ posits a likelihood of the form
\begin{equation}
\mathcal{L}=\prod_{j=1}^{N}\frac{\mathrm{exp}\{-v_{j}^{2}/[2(\sigma_{\rm{GC}}^{2}+\epsilon_{j}^{2})]\}}
{[2\pi(\sigma_{\rm{GC}}^{2}+\epsilon_{j}^{2})^{1/2}]}\ 
,
\end{equation}
where the product is over the 4 GCs and we adopt a Gaussian distribution of velocities and errors. The inferred velocity dispersion is $6.84^{+5.16}_{-3.1}$ (resp. $6.84^{+9.66}_{-4.87}$) with 68\% (resp. 95\%) credibility. The resulting masses, computed using the estimator of Walker et al. (2009, hereafter W09) are shown in Figure 1. By comparison, the masses determined from kinematics of its stellar sub-populations \citep{aae13} are displayed by blue crosses, and the stellar mass profile by the orange stripe. The GC kinematic estimate is clearly incapable of placing strong constraints on the dark matter content of Fornax; indeed, within $2\sigma$, the error bars span a mass range of nearly 2 orders of magnitude; from `no dark matter', to a dark matter content that exceeds the stellar mass by a factor of $\sim 40$.

Fornax's virial mass, on the other hand, may be estimated by considering the empirical relation between the number of GCs and halo mass $M_{200}$ \citep{hudson14,harris17}, assuming that this relation can be extrapolated to dwarf galaxies. We re-write eq. (4) of \cite{harris17} as
\begin{equation}
M_{200}=10^{9.6}M_{\odot}\times N_{\rm GC}^{1.1},
\label{eq:hudson}
\end{equation}
with $\sim 0.3$ dex scatter.

With five GCs, Fornax would have $M_{200}\approx2.5\times10^{10}M_{\odot}$; about five times higher than implied by stellar kinematics if one were to simply do an extrapolation from the measured enclosed mass at the half-light radius. Borrowing from previous nomenclature on UDGs \citep{PvD18,beasley16}, the GCs of Fornax would indicate \emph{an overmassive halo lacking dark matter}. Fornax is not the only example in the Local Group. Indeed, a similar line of argument can be made with the few GCs inhabiting dwarf ellipticals \citep{geha2010} in Andromeda such as NGC147 and NGC 185 for which new GCs have recently been followed up \citep[see][ and references therein]{Veljanoski13}. Clearly, masses estimated from GC abundances are merely indicative in this regime, for which realistic uncertainties are difficult to assign.
\begin{figure}
\includegraphics[width=0.5\textwidth,trim=26mm 200mm 200mm 40mm, clip]{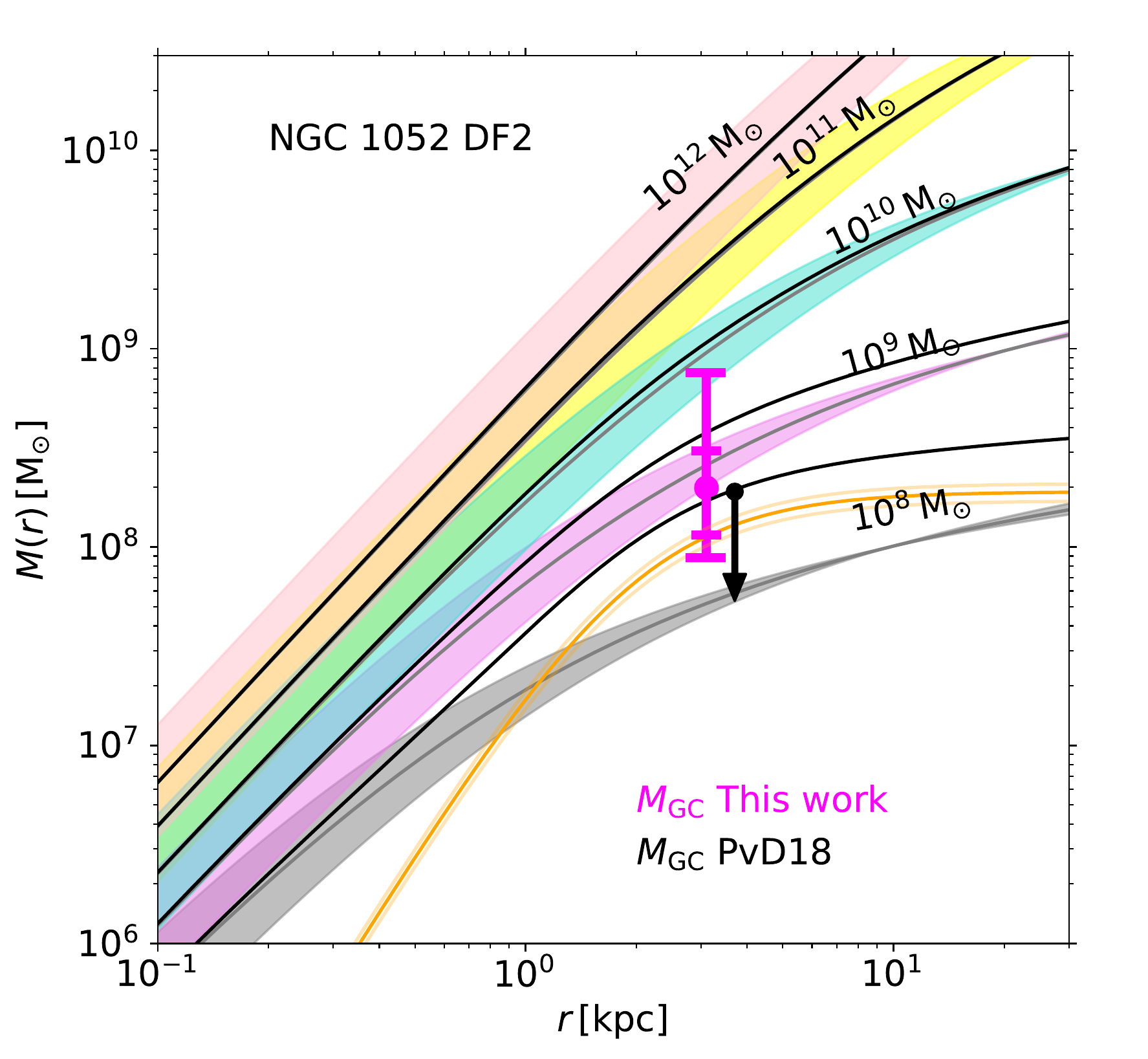}\\
\caption{{
Same as Figure 1, but for the NGC 1052-DF2 ultra-diffuse galaxy. The black symbol is the upper limit on mass from van Dokkum et al. (2018)'s interpretation of the velocity dispersion of the 9 globular clusters, while the magenta point is the corresponding mass that we obtain from the velocity dispersion of all 10 globular clusters.
The GC abundance would yield $M_{200}\approx5.5\times10^{10}M_{\odot}.$ We offset the vD18 point at $R=3.5\,\rm{kpc}$ for visual clarity. 
 }}
\label{fig:udgmasses}
\end{figure}

\subsection{De-biasing observational effects}
\begin{table*} 
\centering
\begin{tabular}{c||c|c|c|c|c||}
\hline
 $N\downarrow$ &  $\sigma/\epsilon_{v}=$0.5 & $\sigma/\epsilon_{v}=$1.0 & $\sigma/\epsilon_{v}=$1.5 & $\sigma/\epsilon_{v}=$2.0 & $\sigma/\epsilon_{v}=$2.5\\
\hline
7 & $[1.7\times10^{-4},0.067,3.16]$ & $[0.02,0.81,2.02]$ & $[0.27,0.88,1.73]$ & $[0.41,0.90,1.62]$ & $[0.44,0.89,1.54]$ \\
11 & $[1.6\times10^{-4},0.74,2.90]$ & $[0.18,0.88,1.83]$ & $[0.41,0.92,1.58]$ & $[0.53,0.93,1.48]$ & $[0.54,0.93,1.45]$ \\
21 & $[2\times10^{-4},0.85,2.46]$ & $[0.39,0.92,1.60]$ & $[0.55,0.95,1.43]$ & $[0.65,0.96,1.35]$ & $[0.67,0.96,1.33]$ \\
31 & $[5\times10^{-4},0.88,2.21]$ & $[0.50,0.96,1.51]$ & $[0.64,0.97,1.35]$ & $[0.71,0.97,1.29]$ & $[0.73,0.97,1.27]$ \\
41 & $[0.003,0.93,2.10]$ & $[0.57,0.97,1.43]$ & $[0.68,0.97,1.32]$ & $[0.75,0.98,1.25]$ & $[0.76,0.98,1.24]$ \\
51 & $[0.02,0.92,1.95]$ & $[0.60,0.97,1.38]$ & $[0.72,0.98,1.28]$ & $[0.78,0.98,1.22]$ & $[0.81,0.99,1.23]$ \\
\hline
$N_{GC}\langle\mu^{2}\rangle/\sigma^{2}$ & 5.0 & 2.0 & 1.4 & 1.14 & 1.1 \\
\hline

\end{tabular}
\caption{Effect of sample size (top to bottom) and ratio $\sigma/\epsilon_{v}$ between `true' dispersion and velocity errors on the inferred, squared velocity dispersion. For each combination, we give the 16-th, 50-th and 84-th percentiles of $\sigma_{\rm mod}^{2}/\sigma^{2},$ where $\sigma$ (resp $\sigma_{\rm mod}$) is the `true' (resp. inferred) velocity dispersion. $10^4$ mocks per parameter choice were used, yielding $\approx1\%$ accuracy in the quantiles. NB: The quantiles may be even lower if the systemic velocity is to be fitted from the tracer kinematics -- as the best-fit average minimizes variance. The final line shows the variance in measured average velocity, across mocks, normalized to $\sigma^{2}/N_{\rm{GC}}:$
 it depends strongly on true-sigma over errors, but not on the sample size, and satisfies the asymptotics ($N_{\rm{GC}}\langle\mu^{2}\rangle/\sigma^{2}\rightarrow 1$)
 at small errors.}
\label{tab:sigmas}
\end{table*}

Masses of dwarf galaxies from GC kinematics usually rely on a small number of tracers, with individual velocity errors that are comparable to the dispersion. The example of NGC 1052-DF2 shows that this is a real concern. Following the same procedure as above, we can quantify the effect of sample size ($N$) and ratio of `true' velocity dispersion to velocity errors ($\sigma/\epsilon_{v}$) on the inferred velocity dispersion $\sigma_{\rm mod}$. For each choice of $N$ and $\sigma/\epsilon_{v}$ we draw $10^4$ mocks, infer their most likely  $\sigma_{\rm mod},$ and then compute the 16-th, 50-th and 84-th percentiles of $\sigma_{\rm mod}^{2}$ across the $10^4$ mocks. We consider the squared velocity dispersion because it is directly related to the mass estimator at $\approx r_{\rm {h}}.$ The results are listed in Table 1. Samples with low $N$ ($\leq11$) or low S/N have a general $\approx20\%$ bias and a large scatter, spanning at least one order of magnitude in the inferred mass at $r_{\rm{h}}$. Even with exquisite velocity measurements on its GCs, NGC~1052-DF2 would have a 50\% systematic scatter in its inferred mass at $\approx 3$~kpc. With the velocity errors of the current sample, its mass at $r_{\rm{h}}$ may be overestimated by $\approx1.7$ or underestimated by almost one order of magnitude. These results can also be applied to ultra-faint galaxies in the Local Group, where velocities are only available for handfuls of stars.

One caveat is in order: the mocks were fit by varying only the velocity dispersion as a free parameter, under the hypothesis that the systemic (average) velocity is determined independently e.g. through starlight spectroscopy of the host. However, this is not always the case, and if the average velocity is another free parameter this has the effect of further biasing the estimated velocity dispersion towards lower values -- as the best-fit average minimizes variance. Then, the quantiles in Table 1 should be taken as optimistic evaluations of bias, and the true bias may be even worse when the systemic velocity is determined directly from the tracer kinematics.
\subsection{NGC 1052-DF2 Revisited}
From the measured GC velocities and errors, vD18 argue that the velocity dispersion $\sigma_{\rm GC}$ of the population must be smaller than $\approx10~\rm{km\,s^{-1}}$ (their 90 percent confidence upper limit), and at most $\approx14.3~\rm{km\,s^{-1}}$. Admittedly, the presence of an `outlier' with $v=(-39\pm14.3)~\rm{km\,s^{-1}}$ is not completely unexpected in a sample of 10 objects with $10~\rm{km\,s^{-1}}\,<\sigma<\,14.3~\rm{km\,s^{-1}}$. Note that the rms velocity errors are of the order $\sim 9\,\rm{km \,s^{-1}}$.
We have re-evaluated the velocity dispersion using the likelihood of eq.~(1), where the product is over the 10 GCs in the sample and we have adopted Gaussian velocity and error distributions. The interested readers are encouraged to see \cite{martin18} for an independent discussion of kinematics in this system, which accounts for possible contamination, and reports results that are comparable to ours. Our estimation yields a most likely $\sigma_{\rm GC}=9.0~\rm{km\,s^{-1}}$, with $7.2<\sigma_{\rm GC}<12.6 \,\rm{km\,s^{-1}}$ at 68\% and $5.3<\sigma_{\rm GC}<20.3 \,\rm{km\,s^{-1}}$ at 95\% confidence levels for the vD18 dataset.\footnote{ Had vD18 used the gapper estimate of velocity dispersion, which is
considered more accurate than the biweight for samples of 10 objects \citep{beers90},
they would have found a velocity dispersion of 13.9 $\rm{km\, s^{-1}}$ instead of 8.4 $\rm{km \,s^{-1}}$, which would
have changed their conclusions.}

This first result is already at variance with the claim of vD18. However, extra caution is needed, as the velocity errors are comparable to $\sigma_{\rm GC}$ itself.
This tends to bias the inferred $\sigma_{\rm GC}$ towards lower values, as has been long known in studies of dwarf galaxies \citep[e.g. see][]{koposov11,toloba16}.
In order to de-bias the inferred dispersion, we drew $10^4$ mocks of 10~GCs from Gaussians with different `true' velocity dispersions, all with $7.0~\rm{km\,s^{-1}}$ velocity errors\footnote{This error estimate is bracketed by the real-data errors, which range between $3~\rm{km\,s^{-1}}$ and $15~\rm{km\,s^{-1}}$.}.
For each mock, we inferred $\sigma_{\rm GC}$ by exploring the likelihood, and then recorded its average and dispersion over the $10^4$ mocks. This procedure is also repeated, for different sample sizes and errors, in the next subsection.
From this de-biasing, we conclude that $\sigma_{\rm GC}=10^{+3}_{-2}~\rm{km\,s^{-1}}$ at 68\% credibility and $\sigma_{\rm GC}=10^{+10.5}_{-3}~\rm{km\,s^{-1}}$ at 95\% credibility.

The mass of NGC 1052-DF2 at $R_{\rm GC}=(3.1\pm0.3)$~kpc, with the debiased $\sigma_{\rm GC}$ quantiles from above, is shown in Figure 2. With the values above, the GCs give a total $M/L\approx 5$ at the 3D half-mass radius $r_{\rm{h}}=3$~kpc, and a dark halo with $M_{200}$ up to $\approx10^{10}M_{\odot}$ is still compatible with the data at 95\% credibility.
Using the abundance estimator of eq.~(\ref{eq:hudson}), with at least 10 GCs\footnote{We could not find the completeness-corrected number of GCs in the vD18 paper, so we are using only the spectroscopically confirmed ones in this estimate.}, we would obtain $M_{200}>5\times10^{10}M_{\odot}.$ The lessons learnt from Fornax, with a significant underestimate (resp. overestimate) from GC kinematics (resp. abundance), would then suggest that NGC~1052-DF2 can easily have $M_{200}\approx5.0\times10^{9}M_{\odot}.$ We also remark that the stellar mass profile plotted by vD18 seems to follow the (2D) enclosed-luminosity profile, instead of the (3D) mass from the de-projected density, and this artificially exacerbates the issue of stellar-vs-dark mass in this system.

\begin{figure}
\includegraphics[width=0.5\textwidth,trim=0mm 0mm 0mm 0mm, clip]{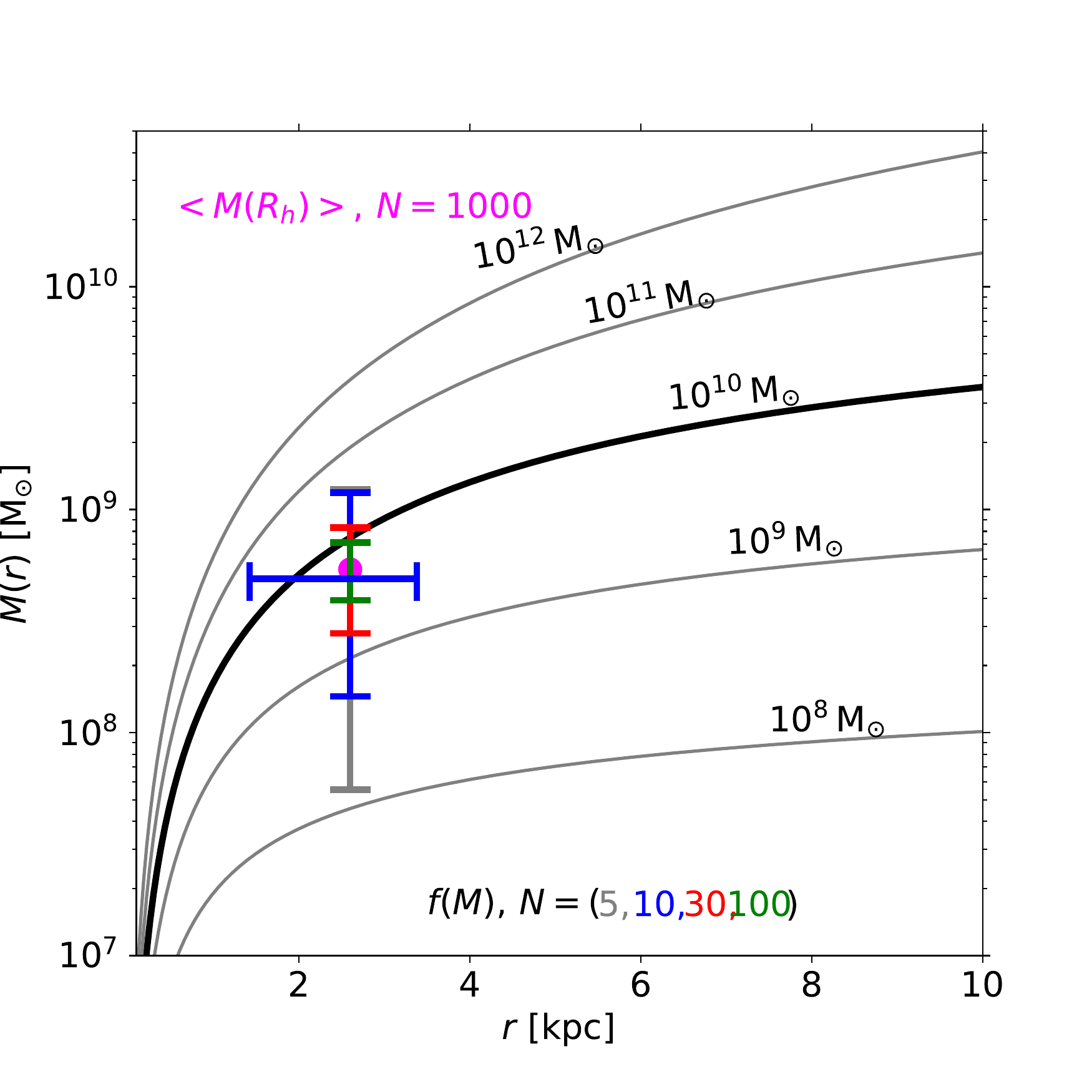}\\
\caption{{Systematics of small samples in dynamical mass estimations using a Plummer tracer population (meant to represent stars or GCs) with $r_{\rm{h}}=2.0 \,\rm{kpc}$ in a halo with $M_{200}=10^{10}\,\rm{M_{\odot}}$ and concentration of $c(M_{200})=12.8$}. The mass profile of the underlying potential is shown in black, as well as the $M(r)$ profiles for halos of $M=[0.01, 0.1, 10, 100]\times10^{10}\,\rm{M_{\odot}}$ in grey. We present the uncertainties within the symmetric 95 \% range in mass estimations for $N=[5,10,30,100]$ tracers in grey, blue, red and green from all the draws. The magenta point marks the median mass estimation from drawing 1000 tracers 1000 times. The blue horizontal line marks the uncertainty (95 \% range) for the mean effective radius derived from using $N=10$ tracers. Systematics (due to sampling) will dominate {\it any} mass-measurements using $N=10$ tracers or fewer.}
\label{fig:masses2}
\end{figure}

\subsection{DF2 re-revisited}
 So far, our findings above can be summarized as follows:

\begin{itemize}

\item With the velocities and distances quoted by vD18a, there is no clear indication that DM is ``lacking'', once a proper analysis of the uncertainties is performed.

\item The mass estimated from GC abundance (11 claimed members) is significantly larger than that from GC kinematics.
\end{itemize}

After April 11th 2018, when this paper was submitted, two updates were made to the data. In a research note, \cite{vD18b} reported deeper data on the GC whose velocity was an outlier and revised it from $-39\, \rm{km\, s^{-1}}$ to $-19 \pm 10\, \rm{km\, s^{-1}}$. In addition, \cite{Trujillo18} revisited the distance estimates to DF2, finding a most likely value of $13\pm 4 \,\rm{kpc}$ instead of the original claim $\sim 19 \pm 1\,\rm{kpc}$. This revised distance would also resolve the claim by \cite{PvD18c} of a striking population of GCs with sizes (resp. luminosities) that are twice (resp. four times) as large as usual. \cite{Trujillo18} also identified another $9$ GC member candidates.

Assuming the distance of van Dokkum, and the revised globular cluster velocity, we find a most-likely $\sigma_{\rm GC}\sim7\,\rm{km\,s^{-1}}$ and median $\sigma_{\rm GC}\sim9\,\rm{km\,s^{-1}}$\footnote{We note that there are MCMC-to-MCMC changes of about $\sim0.5\rm{km\,s^{-1}}$}. Our median de-biased velocity dispersions become $\sigma_{\rm GC}=10.0^{+5}_{-3.5} \,\rm{km\,s^{-1}}$ at 68 \% credibility and $\sigma_{\rm GC}=10.0^{+10.0}_{-5.5} \,\rm{km\,s^{-1}}$ at 95 \% credibility. This would give a mass-to-light ratio at $r_{\rm{GC}}$ of $M/L<6.0$ (90 percent confidence). This still agrees well with our initial calculations and the quoted values of \cite{martin18}.  Assuming that the distance quoted by \cite{Trujillo18} is correct, our mass estimate at $R_{\rm{GC}}\sim 2\,\rm{kpc}$ becomes $M=1.19^{+3.57}_{-0.95}\,10^{8}\,\rm{M_{\odot}}$, in agreement with the 90 percentile value in their Table 4. As it can be appreciated, changing the distance (regardless of its actual value) of DF2 makes no difference to our conclusions.

Aside from the debate on distances, the reason why the uncertainties have not changed much in the velocity dispersion is because the errors on the individual GCs are still {\it large}, regardless of the value of the 10th object. This is much more important than changing the adopted distance which only causes slight changes in the half-light radius and stellar mass of the galaxy. Even when adopting different authors' quoted values for velocities or distances, we are left with upper limits of mass-to-light ratios that are fully consistent with other dwarfs where the evidence for large amounts of dark matter is strong. We note that the estimates quoted by different authors all assume a stellar mass-to-light ratio which is dependent on the adopted IMF, which has typical uncertainties of a factor of 2.

Therefore, no matter what preference one has for the assumed distance of DF2, or even the velocity of the extra GC, we conclude that the dark matter content of DF2 is poorly constrained, either by the available GC kinematics, or by its GC abundance.\footnote{We note that Trujillo detected 9 additional GC candidates, which would exacerbate the inconsistency between the inferred mass from GC kinematics and that from their abundance as discussed above.}, or both. This is not surprising, given our discussion of the Fornax dSph in section 2.1. The important role of GC sample size (not emphasized in other studies of DF2) on inferred masses is further discussed in the two following sub-sections.

\subsection{The risks of small-$N$ systematics using equilibrium distribution functions}
As a final example to illustrate the dangers of small-sample mass estimates, we will use a distribution function (D.F) formalism to generate mocks\footnote{A series of mock observations of spherical systems exist and can be downloaded on the web at the Gaia challenge website: http://astrowiki.ph.surrey.ac.uk/dokuwiki/doku.php?id=tests:sphtri}. 
Beyond the role of the statistical errors in the measurement, we ask ourselves the following question: {\it ``In a world with no measurement errors, how many tracers would I need to make a meaningful mass measurement?''}

We address this by generating the D.F. of an {\it isotropic} tracer population following a Plummer profile in equilibrium inside a spherical Hernquist profile brought to match a NFW profile at the virial radius following \cite{vdmarel12}. We choose $M_{200}=10^{10}\,\rm{M_{\odot}}$, for which a concentration of $c=12.8$ is adopted according to \cite{Correa2015} assuming the cosmological parameters derived by the \cite{planck2016}. We sample the D.F. such that $N=10^{5}$ tracer particles are generated using standard methods \citep[see e.g.][]{Kazantzidis2004, Laporte2015}. It should be noted that for an isotropic distribution function, the distribution of line-of-sight velocities is no longer Gaussian.

We then choose different fiducial numbers of tracers $N=[5,10,30,100,1000]$ to calculate the systematic uncertainties in dynamical mass inferences relying on mass estimators. For this, we choose the estimator of W09, but note that the same exercise with other estimators would produce similar results. The uncertainties are calculated by randomly drawing $N$ tracers 1000 times and calculating $M=580\,(\sigma_{\rm{GC}}/\mathrm{km~s}^{-1})^{2}(r_{\rm{h}}/\mathrm{pc})M_{\odot}$ as in \cite{walker09}, keeping $r_{\rm{h}}=2.6$~kpc fixed. 
The results are shown in Figure 3 against different mass profile curves with $M_{200}=[0.01,0.1,1.0,10,100]\times10^{10}\rm{M_{\odot}}$. As expected \citep{Walker2011}, the W09 method underestimates the enclosed mass as can be seen by the convergent $1000$ tracer median value, albeit by a small amount for our purposes. Interestingly, we see that for systems traced by only 5 to 10 objects the uncertainty in the dynamical mass measurement of a galaxy is always uncertain by at least {\it $\sim 1$ order of magnitude}. This can lead to simultaneously contradictory claims that UDGs bf could live in both MW-mass haloes in some cases
or have very little or no DM within them. 

 Because of the extremely simple model intentionally considered (isotropic, spherical D.F.), these uncertainties should be interpreted as lower limits on the systematic errors (in reality these should be higher). 

It is well known that the measurement of mass can be affected by other sources of systematics. These include i) the anisotropy parameter $\beta$, which is usually not well known for extra-galactic observations which increase the uncertainty in the mass measurement, ii) triaxality, which will introduce severe biases in spherical mass estimators \citep{Laporte2013b} as well as also iii) out-of-equilibrium dynamics. The first two points should be highly relevant to NGC 1052-DF2. Thus the systematics due to small-$N$ samples that we report constitute a lower limit on the true bias introduced by $N$. This, together with the uncertainties from measurement errors on the velocities make the inference of vD18 of a galaxy lacking dark matter all the more inconclusive.

\section{GC/Stellar mass halo connection for the population of Coma UDGs}

Now that we have the shown that NGC 1052-DF2 has a highly uncertain total mass, yet still consistent with it being dark matter dominated, it is useful to see how it compares with the general population of Coma UDGs. Despite the known uncertainties in extrapolating half-light measured masses into virial mass estimates, and our limited knowledge of the luminosity function below LMC-mass like galaxies, we estimate $M_{200}$ assuming that the relation of \cite{harris17} holds for the Coma UDGs \citep{Amorisco2018}. Figure 4 shows where the objects lie in halo mass ($M_{200}$) vs. effective radius and vs. stellar mass. We also show where Fornax would be sitting in those planes if we were to derive its halo mass using its globular cluster count of $N=5$ or extrapolating its measured $M(<r_{\rm{h}})$ to $R_{200}$. We note two aspects from these figures. First, the inferred halo mass ranges of Fornax and NGC~1052-DF2 (from kinematics to $N_{GC}$) do not seem to be in a special category when compared to their other UDG cousins, whether by mass or effective radius. They are in fact consistent with the scatter in the UDG population. Second, we see that Fornax can be considered both as a normal dwarf or as overmassive simply from its GCs, similarly to NGC 1052-DF2 for which the kinematic halo mass extrapolation value is much more uncertain due to low-$N$ systematics. We thus conclude that extra care is needed when interpreting the dark matter content of galaxies through extrapolations.

\begin{figure}

\includegraphics[width=0.5\textwidth, trim=62mm 4mm 90mm 4mm, clip]{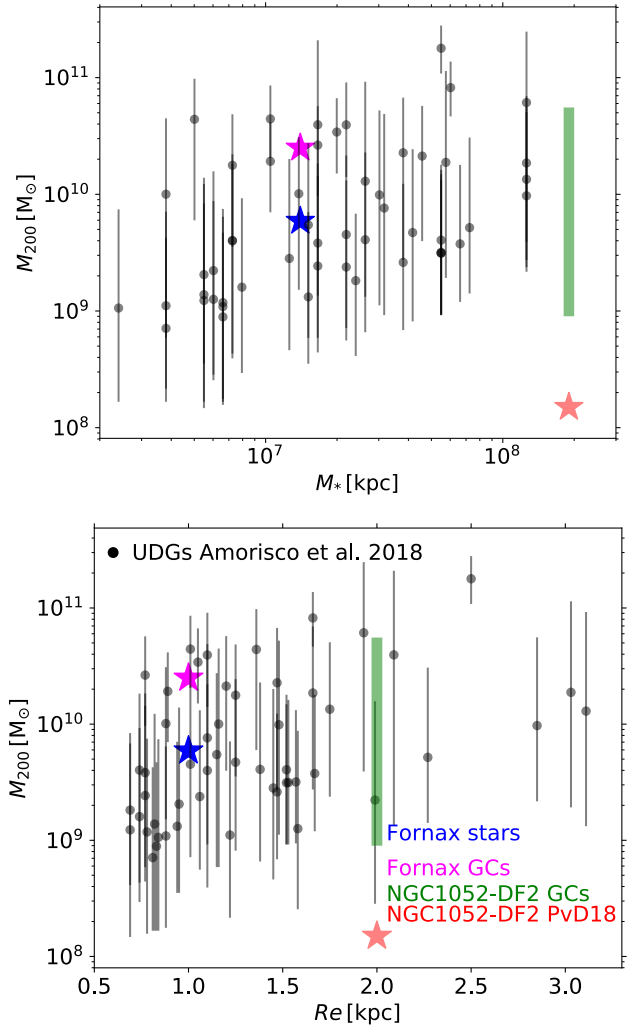}\\
\caption{ Virial mass vs. effective radius (top) and stellar mass (bottom) for the NGC 1052-DF2 ultra-diffuse galaxy (green, from Fig. 2, red from PvD18), the Fornax dwarf spheroidal galaxy (blue using stars, magenta using globular clusters, both from Fig. 1), and
for the ultra-diffuse galaxies in the Coma cluster (black using globular cluster counts). A natural scatter of 1 dex naturally arises because of the expected diversity of halos in LCDM and because of the uncertainties in GC abundances.}
\end{figure}

\section{Discussion}

The `dearth of dark matter' is not a new issue in the literature. Recent claims in this sense were made for massive early-type galaxies \citep[e.g.][]{Romanowsky03}, but were soon recognized as the effect of radial velocity anisotropy \citep{dekel05} or incomplete aperture corrections \citep{agne14}. Here, we have shown how some claims on the DM content of UDGs are likely biased by the use of small samples and inadequate modeling. The NGC~1052-DF2 might eventually prove to lack DM, but our exploration of model systematics and the remarkable case of the Fornax dSph suggest otherwise. Our findings can be summarized as follows:
\begin{itemize}
\item The estimated masses of Fornax from GC abundances and kinematics can make it `overmassive', `just right' or `lacking dark matter', due to large uncertainties from observations, mass estimators, scatter in the mass-concentration relation and tidal stripping.
\item Our revised velocity dispersion of NGC~1052-DF2 and its uncertainties allow for halo masses $M_{200}\approx10^{9}M_{\odot}$ and up to $M_{200}\approx10^{10}M_{\odot},$ for the same reasons.
\item Even with exquisite velocities on 11 GCs, the mass of NGC~1052-DF2 would be systematically uncertain by $\approx50\%$ at $r_{\rm{h}}\approx3$~kpc, and still by an order of magnitude at $M_{200}$ due to the mass-concentration scatter.
\item Besides the model uncertainties from mass-concentration scatter and extrapolation from $r_{\rm{h}}$ to the virial radius, small samples have half-radius mass estimates that are uncertain by at least one order of magnitude, as our distribution-function experiment shows.
\item With masses determined from their GC abundances and kinematics, the Fornax dSph and NGC~1052-DF2 are not unlike the overall population of UDGs in the Coma cluster for which HST imaging data are available. In fact, their GC abundance situates them at the boundary between `normal' and `overmassive' UDGs.
\end{itemize}

Another issue may be that of `sloshing' of GCs relative to the stellar body of their host, which may be indicative of environmental processes on the UDG.
This would be an interesting aspect to investigate with integral-field spectroscopic observations of starlight, as compared to the GCs, also in other galaxies
that may be more tidally disturbed. However, we can expect that small sample size and large velocity uncertainties may produce spurious `sloshing' signals. For this reason, when generating our mocks we also recorded the average velocity $\mu$ of each GC sample, and computed the variance of $\mu$ across different mocks.
The last line in the bias table shows this, normalized to $\sigma^{2}/N_{\rm{GC}},$ such that for very large samples or very small uncertainties the asymptotics $N_{GC}\langle\mu^{2}\rangle/\sigma^{2}\rightarrow 1$ is satisfied. Quite surprisingly, the ratio does not depend on $N_{\rm{GC}},$ up to $N_{\rm{GC}}=50$~GCs. We then conclude that, with the typical sample sizes of GC systems around UDGs and other dwarfs, the sloshing signal may be completely washed out by the measurement errors, unless highly accurate velocities can be obtained. Still, even with highly reliable velocities, measuring sloshing amplitudes below $\sigma/\sqrt{N_{\rm{GC}}}$ would be challenging (or even pointless).

While UDGs are certainly exciting objects, their masses and dark matter content (or lack thereof) will remain highly uncertain for a long time. Some hope remains in the age of next generation telescopes (e.g. MOSAIC on ELT). For now, our best hopes to study dark matter in dwarf galaxies through {\it stellar dynamics} alone remains in our cosmic backyard, for which the Gaia satellite combined with HST archival data is poised to advance the field of dark matter in dwarf galaxies \citep[see e.g.][for some first reports of 3D motions]{Massari18}.

\section*{Acknowledgments}
We thank the referee Gary Mamon for a thorough and useful report which has improved the content of this paper. AA and CL acknowledge the numerous stimulating conversations on low surface brightness galaxies with Nicola~C. Amorisco over the past 3 years. We also thank N.~C. Amorisco, Matthew G. Walker, Jorge Pe\~narrubia, Remco van der Burg and Simon D.~M. White for their critical reading of the manuscript and Nicolas Martin and Benoit Famaey for submitting our manuscripts on the same day. We thank Michelle Collins, Frank van den Bosch, Carlos Frenk, Avi Loeb, Jerry Ostriker, Raja Guha Thakurta, David Spergel, Wyn Evans and Amina Helmi for their words of encouragement and support. This research was supported in part by the National Science Foundation under Grant No. NSF PHY17-48958. 
\bibliographystyle{mn2e}
\bibliography{master2.bib}{}

\begin{thebibliography}{}

\bibitem[\protect\citeauthoryear{{Agnello} \& {Evans}}{{Agnello} \&
  {Evans}}{2012}]{Agnello2012}
{Agnello} A.,  {Evans} N.~W.,  2012, \apjl, 754, L39

\bibitem[\protect\citeauthoryear{{Agnello}, {Evans} \& {Romanowsky}}{{Agnello}
  et~al.}{2014}]{agne14}
{Agnello} A.,  {Evans} N.~W.,    {Romanowsky} A.~J.,  2014, \mnras, 442, 3284

\bibitem[\protect\citeauthoryear{{Amorisco}, {Agnello} \& {Evans}}{{Amorisco}
  et~al.}{2013}]{aae13}
{Amorisco} N.~C.,  {Agnello} A.,    {Evans} N.~W.,  2013, \mnras, 429, L89

\bibitem[\protect\citeauthoryear{{Amorisco} \& {Loeb}}{{Amorisco} \&
  {Loeb}}{2016}]{amo16}
{Amorisco} N.~C.,  {Loeb} A.,  2016, \mnras, 459, L51

\bibitem[\protect\citeauthoryear{{Amorisco}, {Monachesi}, {Agnello} \&
  {White}}{{Amorisco} et~al.}{2018}]{Amorisco2018}
{Amorisco} N.~C.,  {Monachesi} A.,  {Agnello} A.,    {White} S.~D.~M.,  2018,
  \mnras, 475, 4235

\bibitem[\protect\citeauthoryear{{Beasley}, {Romanowsky}, {Pota}, {Navarro},
  {Martinez Delgado}, {Neyer} \& {Deich}}{{Beasley} et~al.}{2016}]{beasley16}
{Beasley} M.~A.,  {Romanowsky} A.~J.,  {Pota} V.,  {Navarro} I.~M.,  {Martinez
  Delgado} D.,  {Neyer} F.,    {Deich} A.~L.,  2016, \apjl, 819, L20

\bibitem[\protect\citeauthoryear{{Beers}, {Flynn} \& {Gebhardt}}{{Beers}
  et~al.}{1990}]{beers90}
{Beers} T.~C.,  {Flynn} K.,    {Gebhardt} K.,  1990, \aj, 100, 32

\bibitem[\protect\citeauthoryear{{Chan}, {Kere{\v s}}, {Wetzel}, {Hopkins},
  {Faucher-Gigu{\`e}re}, {El-Badry}, {Garrison-Kimmel} \&
  {Boylan-Kolchin}}{{Chan} et~al.}{2018}]{chan17}
{Chan} T.~K.,  {Kere{\v s}} D.,  {Wetzel} A.,  {Hopkins} P.~F.,
  {Faucher-Gigu{\`e}re} C.-A.,  {El-Badry} K.,  {Garrison-Kimmel} S.,
  {Boylan-Kolchin} M.,  2018, \mnras, 478, 906

\bibitem[\protect\citeauthoryear{{Cole}, {Dehnen} \& {Wilkinson}}{{Cole}
  et~al.}{2011}]{Cole2011}
{Cole} D.~R.,  {Dehnen} W.,    {Wilkinson} M.~I.,  2011, \mnras, 416, 1118

\bibitem[\protect\citeauthoryear{{Correa}, {Wyithe}, {Schaye} \&
  {Duffy}}{{Correa} et~al.}{2015}]{Correa2015}
{Correa} C.~A.,  {Wyithe} J.~S.~B.,  {Schaye} J.,    {Duffy} A.~R.,  2015,
  \mnras, 452, 1217

\bibitem[\protect\citeauthoryear{{Dalcanton}, {Spergel}, {Gunn}, {Schmidt} \&
  {Schneider}}{{Dalcanton} et~al.}{1997}]{Dalcanton97}
{Dalcanton} J.~J.,  {Spergel} D.~N.,  {Gunn} J.~E.,  {Schmidt} M.,
  {Schneider} D.~P.,  1997, \aj, 114, 635

\bibitem[\protect\citeauthoryear{{Dalcanton}, {Spergel} \&
  {Summers}}{{Dalcanton} et~al.}{1997}]{Dalcanton1997}
{Dalcanton} J.~J.,  {Spergel} D.~N.,    {Summers} F.~J.,  1997, \apj, 482, 659

\bibitem[\protect\citeauthoryear{{Dekel} \& {Silk}}{{Dekel} \&
  {Silk}}{1986}]{dekel86}
{Dekel} A.,  {Silk} J.,  1986, \apj, 303, 39

\bibitem[\protect\citeauthoryear{{Dekel}, {Stoehr}, {Mamon}, {Cox}, {Novak} \&
  {Primack}}{{Dekel} et~al.}{2005}]{dekel05}
{Dekel} A.,  {Stoehr} F.,  {Mamon} G.~A.,  {Cox} T.~J.,  {Novak} G.~S.,
  {Primack} J.~R.,  2005, \nat, 437, 707

\bibitem[\protect\citeauthoryear{{Di Cintio}, {Brook}, {Dutton}, {Macci{\`o}},
  {Obreja} \& {Dekel}}{{Di Cintio} et~al.}{2017}]{dicintio17}
{Di Cintio} A.,  {Brook} C.~B.,  {Dutton} A.~A.,  {Macci{\`o}} A.~V.,  {Obreja}
  A.,    {Dekel} A.,  2017, \mnras, 466, L1

\bibitem[\protect\citeauthoryear{{Famaey}, {McGaugh} \& {Milgrom}}{{Famaey}
  et~al.}{2018}]{famaey18}
{Famaey} B.,  {McGaugh} S.,    {Milgrom} M.,  2018, \mnras, 480, 473

\bibitem[\protect\citeauthoryear{{Geha}, {van der Marel}, {Guhathakurta},
  {Gilbert}, {Kalirai} \& {Kirby}}{{Geha} et~al.}{2010}]{geha2010}
{Geha} M.,  {van der Marel} R.~P.,  {Guhathakurta} P.,  {Gilbert} K.~M.,
  {Kalirai} J.,    {Kirby} E.~N.,  2010, \apj, 711, 361

\bibitem[\protect\citeauthoryear{{Harris}, {Blakeslee} \& {Harris}}{{Harris}
  et~al.}{2017}]{harris17}
{Harris} W.~E.,  {Blakeslee} J.~P.,    {Harris} G.~L.~H.,  2017, \apj, 836, 67

\bibitem[\protect\citeauthoryear{{Hudson}, {Harris} \& {Harris}}{{Hudson}
  et~al.}{2014}]{hudson14}
{Hudson} M.~J.,  {Harris} G.~L.,    {Harris} W.~E.,  2014, \apjl, 787, L5

\bibitem[\protect\citeauthoryear{{Kazantzidis}, {Magorrian} \&
  {Moore}}{{Kazantzidis} et~al.}{2004}]{Kazantzidis2004}
{Kazantzidis} S.,  {Magorrian} J.,    {Moore} B.,  2004, \apj, 601, 37

\bibitem[\protect\citeauthoryear{{Koposov}, {Gilmore}, {Walker}, {Belokurov},
  {Wyn Evans}, {Fellhauer}, {Gieren} \& {Geisler}}{{Koposov}
  et~al.}{2011}]{koposov11}
{Koposov} S.~E.,  {Gilmore} G.,  {Walker} M.~G.,  {Belokurov} V.,  {Wyn Evans}
  N.,  {Fellhauer} M.,  {Gieren} W.,    {Geisler} D.,  2011, \apj, 736, 146

\bibitem[\protect\citeauthoryear{{Laporte}, {Walker} \&
  {Pe{\~n}arrubia}}{{Laporte} et~al.}{2013}]{Laporte2013b}
{Laporte} C.~F.~P.,  {Walker} M.~G.,    {Pe{\~n}arrubia} J.,  2013, \mnras,
  433, L54

\bibitem[\protect\citeauthoryear{{Laporte} \& {White}}{{Laporte} \&
  {White}}{2015}]{Laporte2015}
{Laporte} C.~F.~P.,  {White} S.~D.~M.,  2015, \mnras, 451, 1177

\bibitem[\protect\citeauthoryear{{Martin}, {Collins}, {Longeard} \&
  {Tollerud}}{{Martin} et~al.}{2018}]{martin18}
{Martin} N.~F.,  {Collins} M.~L.~M.,  {Longeard} N.,    {Tollerud} E.,  2018,
  \apjl, 859, L5

\bibitem[\protect\citeauthoryear{{Martin}, {Ibata}, {Lewis}, {McConnachie},
  {Babul}, {Bate}, {Bernard}, {Chapman}, {Collins}, {Conn}, {Crnojevi{\'c}},
  {Fardal}, {Ferguson}, {Irwin}, {Mackey}, {McMonigal}, {Navarro} \&
  {Rich}}{{Martin} et~al.}{2016}]{Martin2016}
{Martin} N.~F.,  {Ibata} R.~A.,  {Lewis} G.~F.,  {McConnachie} A.,  {Babul} A.,
   {Bate} N.~F.,  {Bernard} E.,  {Chapman} S.~C.,  {Collins} M.~M.~L.,  {Conn}
  A.~R.,  {Crnojevi{\'c}} D.,  {Fardal} M.~A.,  {Ferguson} A.~M.~N.,  {Irwin}
  M.,  {Mackey} A.~D.,  {McMonigal} B.,  {Navarro} J.~F.,    {Rich} R.~M.,
  2016, \apj, 833, 167

\bibitem[\protect\citeauthoryear{{Mart{\'{\i}}nez-Delgado}, {Pe{\~n}arrubia},
  {Gabany}, {Trujillo}, {Majewski} \& {Pohlen}}{{Mart{\'{\i}}nez-Delgado}
  et~al.}{2008}]{delgado08}
{Mart{\'{\i}}nez-Delgado} D.,  {Pe{\~n}arrubia} J.,  {Gabany} R.~J.,
  {Trujillo} I.,  {Majewski} S.~R.,    {Pohlen} M.,  2008, \apj, 689, 184

\bibitem[\protect\citeauthoryear{{Massari}, {Breddels}, {Helmi}, {Posti},
  {Brown} \& {Tolstoy}}{{Massari} et~al.}{2018}]{Massari18}
{Massari} D.,  {Breddels} M.~A.,  {Helmi} A.,  {Posti} L.,  {Brown} A.~G.~A.,
   {Tolstoy} E.,  2018, Nature Astronomy, 2, 156

\bibitem[\protect\citeauthoryear{{Navarro} \& {White}}{{Navarro} \&
  {White}}{1994}]{Navarro94}
{Navarro} J.~F.,  {White} S.~D.~M.,  1994, \mnras, 267, 401

\bibitem[\protect\citeauthoryear{{Planck Collaboration}, {Ade}, {Aghanim},
  {Arnaud}, {Ashdown}, {Aumont}, {Baccigalupi}, {Banday}, {Barreiro},
  {Bartlett} \& et al.}{{Planck Collaboration} et~al.}{2016}]{planck2016}
{Planck Collaboration} {Ade} P.~A.~R.,  {Aghanim} N.,  {Arnaud} M.,  {Ashdown}
  M.,  {Aumont} J.,  {Baccigalupi} C.,  {Banday} A.~J.,  {Barreiro} R.~B.,
  {Bartlett} J.~G.,    et al. 2016, \aap, 594, A13

\bibitem[\protect\citeauthoryear{{Rom{\'a}n} \& {Trujillo}}{{Rom{\'a}n} \&
  {Trujillo}}{2017}]{Roman17}
{Rom{\'a}n} J.,  {Trujillo} I.,  2017, \mnras, 468, 4039

\bibitem[\protect\citeauthoryear{{Romanowsky}, {Douglas}, {Arnaboldi},
  {Kuijken}, {Merrifield}, {Napolitano}, {Capaccioli} \&
  {Freeman}}{{Romanowsky} et~al.}{2003}]{Romanowsky03}
{Romanowsky} A.~J.,  {Douglas} N.~G.,  {Arnaboldi} M.,  {Kuijken} K.,
  {Merrifield} M.~R.,  {Napolitano} N.~R.,  {Capaccioli} M.,    {Freeman}
  K.~C.,  2003, Science, 301, 1696

\bibitem[\protect\citeauthoryear{{Sandage} \& {Binggeli}}{{Sandage} \&
  {Binggeli}}{1984}]{Sandage1984}
{Sandage} A.,  {Binggeli} B.,  1984, \aj, 89, 919

\bibitem[\protect\citeauthoryear{{Shang}, {Zheng}, {Brinks}, {Chen}, {Burstein}
  \& {Su}}{{Shang} et~al.}{1998}]{Shang98}
{Shang} Z.,  {Zheng} Z.,  {Brinks} E.,  {Chen} J.,  {Burstein} D.,    {Su} H.,
  1998, \apjl, 504, L23

\bibitem[\protect\citeauthoryear{{Toloba}, {Li}, {Guhathakurta}, {Peng},
  {Ferrarese}, {C{\^o}t{\'e}}, {Emsellem}, {Gwyn}, {Zhang}, {Boselli},
  {Cuillandre}, {Jordan} \& {Liu}}{{Toloba} et~al.}{2016}]{toloba16}
{Toloba} E.,  {Li} B.,  {Guhathakurta} P.,  {Peng} E.~W.,  {Ferrarese} L.,
  {C{\^o}t{\'e}} P.,  {Emsellem} E.,  {Gwyn} S.,  {Zhang} H.,  {Boselli} A.,
  {Cuillandre} J.-C.,  {Jordan} A.,    {Liu} C.,  2016, \apj, 822, 51

\bibitem[\protect\citeauthoryear{{Toloba}, {Lim}, {Peng}, {Sales},
  {Guhathakurta}, {Mihos}, {C{\^o}t{\'e}}, {Boselli}, {Cuillandre},
  {Ferrarese}, {Gwyn}, {Lan{\c c}on}, {Mu{\~n}oz} \& {Puzia}}{{Toloba}
  et~al.}{2018}]{toloba18}
{Toloba} E.,  {Lim} S.,  {Peng} E.,  {Sales} L.~V.,  {Guhathakurta} P.,
  {Mihos} J.~C.,  {C{\^o}t{\'e}} P.,  {Boselli} A.,  {Cuillandre} J.-C.,
  {Ferrarese} L.,  {Gwyn} S.,  {Lan{\c c}on} A.,  {Mu{\~n}oz} R.,    {Puzia}
  T.,  2018, \apjl, 856, L31

\bibitem[\protect\citeauthoryear{{Torrealba}, {Koposov}, {Belokurov} \&
  {Irwin}}{{Torrealba} et~al.}{2016}]{Torrealba16}
{Torrealba} G.,  {Koposov} S.~E.,  {Belokurov} V.,    {Irwin} M.,  2016,
  \mnras, 459, 2370

\bibitem[\protect\citeauthoryear{{Trujillo}, {Beasley}, {Borlaff}, {Carrasco},
  {Di Cintio}, {Filho}, {Monelli}, {Montes}, {Roman}, {Ruiz-Lara}, {Sanchez
  Almeida}, {Valls-Gabaud} \& {Vazdekis}}{{Trujillo} et~al.}{2018}]{Trujillo18}
{Trujillo} I.,  {Beasley} M.~A.,  {Borlaff} A.,  {Carrasco} E.~R.,  {Di Cintio}
  A.,  {Filho} M.,  {Monelli} M.,  {Montes} M.,  {Roman} J.,  {Ruiz-Lara} T.,
  {Sanchez Almeida} J.,  {Valls-Gabaud} D.,    {Vazdekis} A.,  2018, ArXiv
  e-prints

\bibitem[\protect\citeauthoryear{{Turner} \& {et al.}}{{Turner} \& {et
  al.}}{1993}]{Turner1993}
{Turner} J.~A.,  {et al.} 1993, \mnras, 261, 39

\bibitem[\protect\citeauthoryear{{van der Burg}, {Hoekstra}, {Muzzin},
  {Sif{\'o}n}, {Viola}, {Bremer}, {Brough} \& {Driver}}{{van der Burg}
  et~al.}{2017}]{vdb17}
{van der Burg} R.~F.~J.,  {Hoekstra} H.,  {Muzzin} A.,  {Sif{\'o}n} C.,
  {Viola} M.,  {Bremer} M.~N.,  {Brough} S.,    {Driver} S.~P.,  2017, \aap,
  607, A79

\bibitem[\protect\citeauthoryear{{van der Marel} \& {et al.}}{{van der Marel}
  \& {et al.}}{2012}]{vdmarel12}
{van der Marel} R.~P.,  {et al.} 2012, \apj, 753, 8

\bibitem[\protect\citeauthoryear{{van Dokkum}, {Abraham}, {Brodie}, {Conroy},
  {Danieli}, {Merritt}, {Mowla}, {Romanowsky} \& {Zhang}}{{van Dokkum}
  et~al.}{2016}]{PvD16}
{van Dokkum} P.,  {Abraham} R.,  {Brodie} J.,  {Conroy} C.,  {Danieli} S.,
  {Merritt} A.,  {Mowla} L.,  {Romanowsky} A.,    {Zhang} J.,  2016, \apjl,
  828, L6

\bibitem[\protect\citeauthoryear{{van Dokkum}, {Cohen}, {Danieli}, {Kruijssen},
  {Romanowsky}, {Merritt}, {Abraham}, {Brodie}, {Conroy}, {Lokhorst}, {Mowla},
  {O Sullivan} \& {Zhang}}{{van Dokkum} et~al.}{2018}]{PvD18c}
{van Dokkum} P.,  {Cohen} Y.,  {Danieli} S.,  {Kruijssen} J.~M.~D.,
  {Romanowsky} A.~J.,  {Merritt} A.,  {Abraham} R.,  {Brodie} J.,  {Conroy} C.,
   {Lokhorst} D.,  {Mowla} L.,  {O Sullivan} E.,    {Zhang} J.,  2018, \apjl,
  856, L30

\bibitem[\protect\citeauthoryear{{van Dokkum}, {Cohen}, {Danieli},
  {Romanowsky}, {Abraham}, {Brodie}, {Conroy}, {Kruijssen}, {Lokhorst},
  {Merritt}, {Mowla} \& {Zhang}}{{van Dokkum} et~al.}{2018}]{vD18b}
{van Dokkum} P.,  {Cohen} Y.,  {Danieli} S.,  {Romanowsky} A.,  {Abraham} R.,
  {Brodie} J.,  {Conroy} C.,  {Kruijssen} J.~M.~D.,  {Lokhorst} D.,  {Merritt}
  A.,  {Mowla} L.,    {Zhang} J.,  2018, Research Notes of the American
  Astronomical Society, 2, 54

\bibitem[\protect\citeauthoryear{{van Dokkum}, {Danieli}, {Cohen}, {Merritt},
  {Romanowsky}, {Abraham}, {Brodie}, {Conroy}, {Lokhorst}, {Mowla},
  {O'Sullivan} \& {Zhang}}{{van Dokkum} et~al.}{2018}]{PvD18}
{van Dokkum} P.,  {Danieli} S.,  {Cohen} Y.,  {Merritt} A.,  {Romanowsky}
  A.~J.,  {Abraham} R.,  {Brodie} J.,  {Conroy} C.,  {Lokhorst} D.,  {Mowla}
  L.,  {O'Sullivan} E.,    {Zhang} J.,  2018, \nat, 555, 629

\bibitem[\protect\citeauthoryear{{van Dokkum}, {Abraham} \& {Merritt}}{{van
  Dokkum} et~al.}{2014}]{Pvd14}
{van Dokkum} P.~G.,  {Abraham} R.,    {Merritt} A.,  2014, \apjl, 782, L24

\bibitem[\protect\citeauthoryear{{Veljanoski}, {Ferguson}, {Huxor}, {Mackey},
  {Fishlock}, {Irwin} \& {Tanvir}}{{Veljanoski} et~al.}{2013}]{Veljanoski13}
{Veljanoski} J.,  {Ferguson} A.~M.~N.,  {Huxor} A.~P.,  {Mackey} A.~D.,
  {Fishlock} C.~K.,  {Irwin} M.~J.,    {Tanvir} N.,  2013, \mnras, 435, 3654

\bibitem[\protect\citeauthoryear{{Walker}, {Mateo}, {Olszewski},
  {Pe{\~n}arrubia}, {Wyn Evans} \& {Gilmore}}{{Walker} et~al.}{2009}]{walker09}
{Walker} M.~G.,  {Mateo} M.,  {Olszewski} E.~W.,  {Pe{\~n}arrubia} J.,  {Wyn
  Evans} N.,    {Gilmore} G.,  2009, \apj, 704, 1274

\bibitem[\protect\citeauthoryear{{Walker} \& {Pe{\~n}arrubia}}{{Walker} \&
  {Pe{\~n}arrubia}}{2011}]{Walker2011}
{Walker} M.~G.,  {Pe{\~n}arrubia} J.,  2011, \apj, 742, 20

\bibitem[\protect\citeauthoryear{{Wolf}, {Martinez}, {Bullock}, {Kaplinghat},
  {Geha}, {Mu{\~n}oz}, {Simon} \& {Avedo}}{{Wolf} et~al.}{2010}]{Wolf2010}
{Wolf} J.,  {Martinez} G.~D.,  {Bullock} J.~S.,  {Kaplinghat} M.,  {Geha} M.,
  {Mu{\~n}oz} R.~R.,  {Simon} J.~D.,    {Avedo} F.~F.,  2010, \mnras, 406, 1220

\end{thebibliography}

\label{lastpage}
\end{document}